\documentclass[twocolumn]{aastex61}
\pdfoutput=1 
\usepackage{amsmath,amstext,amssymb}
\usepackage[T1]{fontenc}
\usepackage{apjfonts} 
\usepackage[figure,figure*]{hypcap}
\usepackage{natbib}
\usepackage{flushend}
\usepackage{bm}

\newcommand\kms{\ifmmode{\rm km\thinspace s^{-1}}\else km\thinspace s$^{-1}$\fi}
\newcommand\ms{\ifmmode{\rm m\thinspace s^{-1}}\else m\thinspace s$^{-1}$\fi}
\newcommand\msun{\ifmmode{M_{\odot}}\else $M_{\odot}$\fi}
\newcommand\rsun{\ifmmode{R_{\odot}}\else $R_{\odot}$\fi}
\newcommand\lsun{\ifmmode{L_{\odot}}\else $L_{\odot}$\fi}
\newcommand\mstar{\ifmmode{M_{\star}}\else $M_{\star}$\fi}
\newcommand\rstar{\ifmmode{R_{\star}}\else $R_{\star}$\fi}
\newcommand\mjup{\ifmmode{M_{\rm Jup}}\else $M_{\rm Jup}$\fi}
\newcommand\rjup{\ifmmode{R_{\rm Jup}}\else $R_{\rm Jup}$\fi}
\newcommand\mearth{\ifmmode{M_\oplus}\else $M_\oplus$\fi}
\newcommand\rearth{\ifmmode{R_\oplus}\else $R_\oplus$\fi}
\newcommand\mpl{\ifmmode{M_{\rm P}}\else $M_{\rm P}$\fi}
\newcommand\rp{\ifmmode{R_{\rm P}}\else $R_{\rm P}$\fi}
\newcommand\vsini{\ifmmode{v\sin{i_\star}}\else $v\sin{i_\star}$\fi}
\newcommand\sini{\ifmmode{\sin{i_\star}}\else $\sin{i_\star}$\fi}
\newcommand\prot{\ifmmode{P_{\rm rot}}\else $P_{\rm rot}$\fi}
\newcommand\logg{\ifmmode{\log{g}}\else $\log{g}$\fi}
\newcommand\teff{\ifmmode{T_{\rm eff}}\else $T_{\rm eff}$\fi}
\newcommand\kep{{\em Kepler}}
\def\vmac{$v_{\rm mac}$}
\newcommand{\feh}{\mbox{$\rm{[Fe/H]}$}}

\newcommand\mysim{\mathord{\sim}}
\newcommand\dg{\ensuremath{^\circ}}
\def\lsim{\mathrel{\rlap{\lower4pt\hbox{\hskip1pt$\sim$}}
    \raise1pt\hbox{$<$}}}
\def\gsim{\mathrel{\rlap{\lower4pt\hbox{\hskip1pt$\sim$}}
    \raise1pt\hbox{$>$}}}
\newcommand{\reffig}[1]{Fig.~\ref{fig:#1}}
\newcommand{\reffigl}[1]{Figure~\ref{fig:#1}}

\newcommand{\refsec}[1]{\mbox{\S\ \ref{sec:#1}}}
\newcommand{\refsecl}[1]{\mbox{Section \ref{sec:#1}}}
\newcommand{\refsecs}[2]{\mbox{Sections \ref{sec:#1} and \ref{sec:#2}}}

\newcommand{\reftabl}[1]{Table~\ref{tab:#1}}

\newcommand{\refeql}[1]{Equation~\ref{eq:#1}}

\makeatletter
\newcommand*{\hyperlinkcite}[1]{\hyper@link{cite}{cite.#1}}
\makeatother
\newcommand*{\sull}{\hyperlinkcite{sullivan:2015}{S15}}

\journalinfo{Published in ApJ December 16, 2016}
\shorttitle{Exoplanet Host Star Obliquities}
\shortauthors{Quinn \& White}

\begin{document}

\title{Obliquities of Exoplanet Host Stars from Precise Distances \\and Stellar Angular Diameters}

\author{Samuel N. Quinn}
\affil{Harvard-Smithsonian Center for Astrophysics, 60 Garden Street, Cambridge, MA 02138, USA; \href{mailto:squinn@cfa.harvard.edu}{squinn@cfa.harvard.edu}}
\author{Russel J. White}
\affil{Department of Physics \& Astronomy, Georgia State University, 25 Park Place Suite 605, Atlanta, GA, 30303, USA}

\begin{abstract}
The next generation of exoplanet space photometry missions proposed by
both NASA and ESA promise to discover small transiting planets around
the nearest and brightest main-sequence stars. The physical and
rotational properties of these stars, in conjunction with
Gaia-precision distances, can be used to determine the inclination of
the stellar rotation axis. Given edge-on orbital paths for transiting
planets, stellar inclinations can be interpreted as obliquities
projected into the line of sight, which can be used to more clearly
reveal the system architectures of small planets and the factors that
drive their orbital evolution. To demonstrate the method, we use a
sample of simulated target stars for the NASA Transiting Exoplanet
Survey Satellite (TESS) mission. Based on predicted characteristics of
these stars and likely measurement uncertainties, we show that the
expected TESS discoveries will allow us to finely differentiate the
true underlying obliquity distribution. Under conservative assumptions
in our illustrative example---in which the true distribution is
assumed to contain systems drawn from both well-aligned and isotropic
distributions (e.g., due to multiple migration channels)---the correct
fractions can be determined to within $0.15$, thus enabling
constraints on the evolutionary processes that shape system
architectures. Moreover, because of the excellent astrometric precision 
expected from Gaia, this technique will also be applicable to the large 
number of planets already discovered by {\em Kepler} orbiting much more 
distant stars.
\end{abstract}

\keywords{planets and satellites: dynamical
evolution and stability -- planets and satellites: formation --
planets and satellites: terrestrial planets}

\section{Introduction}

The stellar obliquity of a planetary system ($\psi$, the angle between
the orbital momentum vector and the stellar rotational momentum
vector) is shaped by a combination of several factors,
including the degree of primordial misalignment between the stellar
spin and the protoplanetary disk out of which the planet forms,
dynamical interactions with other planets or stars, and tidal or
magnetic interaction with the host star \citep[see review
by][]{winn:2015}. The obliquity can thus potentially provide valuable
observational constraints on the planetary environment at formation,
the architecture of planetary systems, and the processes that most
often drive changes in planetary systems.

While direct measurements of the obliquity are in practice difficult,
observational techniques exist that allow for precise measurements of
two different projections of the obliquity. The first of these is the
projection of obliquity on the plane of the sky, commonly referred to
as $\lambda$ \citep[see Figure 1 from][]{schlaufman:2010}; this
measures how azimuthally aligned the vectors are, but does not
constrain the inclination along our line of sight. Techniques that
enable measurements of $\lambda$\ include the Rossiter--McLaughlin
effect \citep{rossiter:1924,mclaughlin:1924,queloz:2000,winn:2007},
Doppler tomography \citep[e.g.,][]{cameron:2010}, and planets that
transit spotted stellar surfaces
\citep[e.g.,][]{sanchis:2011,nutzman:2011,desert:2011} or
gravity-darkened stars \citep[e.g.,][]{barnes:2011b}. In special cases,
the Doppler beaming "photometric Rossiter--McLaughlin" effect has been
used \citep{groot:2012,shporer:2012}, but this is not widely
applicable.

Recent use of these techniques for hot Jupiter systems provides good
examples of the potential for projected obliquity measurements to
place constraints on the dynamical evolution of planetary systems. One
seminal result is that hot Jupiters are found with a wide range of
obliquities, including orbits that are highly inclined, polar
\citep[e.g., WASP-7b;][]{albrecht:2012a}, or even retrograde
\citep[e.g., WASP-15b;][]{triaud:2010}, which is interpreted as
evidence that inward migration is often driven by multi-body
interactions (rather than through the gas disk). Moreover, hot
Jupiters orbiting hot stars (${\gtrsim}6250$\,K) often have high
obliquities, whereas cool stars (${\lesssim}6250$\,K) tend to be
well-aligned \citep[e.g.,][]{winn:2010a}, which suggests a difference
in formation, migration, or dynamical interaction between the two
groups. Following a more physical motivation, the sample can as easily
be divided into stars with short and long timescales for tidal
dissipation \citep{albrecht:2012b} or stars with strong versus weak
magnetic braking \citep{dawson:2014a}, perhaps indicating that all hot
Jupiters initially have a range of obliquities, but cool stars are
able to realign more efficiently. This dichotomy could also be
explained by realignment between the star and an initially misaligned
planet-forming disk \citep{spalding:2015}, as low-mass T-Tauri stars
have much stronger magnetic fields than their high-mass counterparts.

While these observations are beginning to place helpful constraints on
the dynamical interactions that are important for shaping systems of
gas giant planets, relatively little is known about the orbital
histories of small planets, despite their relative abundance
\citep[e.g.,][]{swift:2013,fressin:2013,dressing:2013,petigura:2013,muirhead:2015,ballard:2016}.
This is because the techniques that can successfully measure $\lambda$
are currently extremely difficult or infeasible for small planets due
to the slight flux they occult. Nevertheless, population analyses of
the stars that host small planets are providing indirect evidence of
typical spin-axis orientations. For example, \citet{mazeh:2015} found
that the average amplitude of photometric rotational modulation of
cool \kep\ transiting planet hosts is larger than that for stars with no
transits. They conclude that this is the result of typically edge-on
equators (well-aligned with the orbital plane) for cool hosts
(${<}6000$\,K), and they further show that this result extends to
orbital periods of at least $50$\ days. While extension to such long
periods would be unexpected if alignment were due solely to tidal
realignment, as has been proposed for hot Jupiters, \citet{li:2016}
expand upon this to show that the photometric amplitude is larger for
shorter orbital periods, seemingly corroborating the idea that tides may 
play a role.

It is also now well known that systems of small planets tend to be
mutually well-aligned \citep[i.e.,
  coplanar;][]{fabrycky:2014,fang:2012}, which is suggestive of a
calmer dynamical history for small planets compared to hot
Jupiters---small planets inside the ice line may have predominantly
experienced disk migration, or very little migration at all if many of
them form in situ \citep[e.g.,][]{hansen:2012,chiang:2013}. However,
measurement of the inclination of the orbital planes with respect to
the star is in most cases unknown. While one would generally expect
spin-orbit alignment for systems with mutual orbital alignment, the
disk itself could become misaligned as a consequence of the chaotic
environment of star formation
\citep[e.g.,][]{bate:2010,thies:2011,fielding:2015}, or through
star--disk interactions \citep{lai:2011,batygin:2013}. Gravitational
perturbation from outer companions may also tilt orbital planes while
the inner planets remain mutually coplanar. This has been proposed as
the mechanism responsible for Kepler-56, a system with spin-orbit
misalignment, but two coplanar (transiting) planets and an outer
massive companion \citep{huber:2013a}. Direct measures of the
obliquities of small planets could elucidate the formation conditions
and processes that drive spin-orbit misalignment for small planets.

A second projection of the obliquity---its projection onto our line of
sight---is equivalent to the inclination of the stellar spin axis
$i_\star$ since the presence of transits implies that $i_{\rm orb}\mysim 
90\dg$; this measures how similar the inclinations are,
but does not constrain their relative azimuthal orientation. The
advantage to using measurements of $i_\star$ over measurements of
$\lambda$\ is that it depends only upon the properties of the star;
$i_\star$\ can be measured equally well for stars that host either
small or large planets.

For example, \citet{dumusque:2014} demonstrate that a sophisticated
star spot and stellar activity model can constrain $i_\star$, but this
is relatively imprecise and requires significant spectroscopic
follow-up. The stellar inclination can also be measured via
asteroseismic modeling of the rotational splitting of oscillation
modes in the stellar interior
\citep[e.g.,][]{chaplin:2013,huber:2013a,quinn:2015}. While this does
not depend on the size of the planet, it unfortunately requires long
time span, precise photometry and strong stellar oscillations; even
among the \kep\ data set, only a small fraction of main sequence stars
is amenable to this technique. A classic and perhaps more robust
method for estimating $i_\star$\ relies on using the rotation period,
\prot, the stellar radius, $R_\star$, and the projected rotational
velocity, \vsini.
\begin{equation} \label{eq:1}
  \sin{i_\star} = \frac{v \sin{i_\star} P_{\rm rot}}{2 \pi R_\star}.
\end{equation}
Sometimes called the \vsini\ method, we refer to this as ``inclination 
from projected rotation'', or IPR, and use this initialism throughout
for brevity. The main drawback of IPR is that even when precise
measurements of \prot\ are acquired (e.g., from the same photometry
used to discover a transiting planet), only very rough constraints on
$i_\star$\ (several tens of degrees) have generally been possible due
to significant uncertainties associated with estimates of $R_\star$
and, for slow rotators, with \vsini. Furthermore, because the sine
function flattens near $90\dg$, obliquities for
well-aligned systems will be imprecise, even for high precision
measurements. IPR is therefore most effective for identifying
misaligned orbits. Nevertheless, the growing population of these
systems is beginning to provide statistical constraints on the
obliquities of hosts of both large and small exoplanets.

\citet{schlaufman:2010} employed IPR for hot Jupiters in a statistical
sense, adopting predicted values for \prot\ and model-dependent radii
to show agreement with emerging overall obliquity trends, and to
identify anomalously slow rotators likely indicative of
misalignment. Others
\citep{hirano:2012,hirano:2014,walkowicz:2013,morton:2014} have
applied IPR to \kep\ systems, to interesting
effect. \citet{hirano:2014} report a statistical tendency for
alignment among $25$\ systems, and identify several that may be
misaligned. Using a sample of $70$\ systems and a more sophisticated
statistical framework, \citet{morton:2014} characterize the obliquity
distribution according to the Fisher distribution parameter $\kappa$,
and find a hint that multiple-transiting systems have lower
obliquities than systems with a single transiting planet. They suggest
this may be evidence that the singles are actually part of a
population of ``dynamically hot'' multiple planet systems that have
large mutual inclinations, distinct from the remarkably flat systems
revealed by the presence of multiple transiting
planets. \citet{li:2016} also find evidence (via the enhanced
photometric amplitudes of \kep\ hosts with multiple transiting
planets) for these distinct populations. A more detailed
characterization of the obliquity distribution should shed light on
the formation and evolutionary processes that have shaped these
planetary systems.

In this paper, we highlight a confluence of new observational surveys
and facilities that will enable more precise measurements of
$i_\star$\ for large samples of nearby transiting exoplanet systems.
Paramount to this is the success of proposed photometric survey
satellite missions designed to identify small transiting exoplanets
orbiting the nearest and brightest main sequence stars. Proposed for
launch in 2017 is an ESA S-class mission, the CHaracterizing ExOPlanet
Satellite \citep[CHEOPS;][]{broeg:2013}. CHEOPS will observe $\sim 
500$ nearby stars with previously identified transiting planets via
radial velocity and/or photometric measurements. It will thereby
increase the total number of known nearby transiting planet systems.
Also scheduled for launch in 2017 is a NASA Explorer mission, the
Transiting Exoplanet Survey Satellite \citep[TESS;][]{ricker:2014}.
TESS will monitor at least $\mysim 200,\!000$\ nearby main sequence
stars to identify new transiting exoplanets; most exoplanet
discoveries will be short period ($< 20$ days) because of its
observing strategy. Within the next decade (circa 2022--2024) ESA
proposes to launch the Planetary Transits and Oscillations of stars
mission \citep[PLATO 2.0;][]{rauer:2014}. PLATO 2.0 will survey up to
1 million nearby main sequence stars with an observing strategy that
will identify and characterize terrestrial planets at orbital
distances that extend to the habitable zone for Sun-like stars. The
observational goals and thus observing strategy of PLATO 2.0 may
evolve depending upon the discoveries of CHEOPS and TESS
\citep{rauer:2014}.

Likely prior to any of these transit missions being launched, precise
distances for all of these stars will be available from ESA's Gaia
mission, which is currently measuring trigonometric parallaxes with a
predicted precision of $\lesssim 10\ \mu$as
\citep{bruijne:2012,bruijne:2014}. Once combined with broadband
photometry now available for the majority of nearby stars
\citep[e.g.,][]{stassun:2014}, these distances will enable much more
accurate determinations of bolometric fluxes and estimated radii. The
proximity of the bright stars targeted in these missions will also
permit long baseline optical/infrared interferometry to directly
measure the sizes of many, from which radii can very accurately be
determined \citep[e.g.,][]{braun:2014}, or from which relations
that improve radius estimates can be extracted
\citep[e.g.,][]{boyajian:2014}. The dramatically improved precision of
these measurements of nearby stars will lead to a corresponding
improvement in the efficacy of IPR and, consequently, the resulting
constraints on processes that shape orbital architectures.

In \refsecl{method}, we describe IPR using observed stellar angular
diameters and distances, rather than radius estimates as has been
employed previously. In \refsecl{sample} we define the adopted
simulated stellar population, including the rotational properties of
this population. We discuss appropriate observational uncertainties
and simulate the observations in \refsecl{errors}, and we present
results---obliquities of individual systems and constraints on
underlying obliquity distributions---in \refsecl{results}. Finally, we
consider additional applications and offer further discussion in
\refsecs{apps}{summary}.

\section{The Method}
\label{sec:method}

As seen in \refeql{1}, the idea behind IPR is mathematically
straightforward: the sine of the stellar inclination can be calculated
by dividing the projected rotational velocity (\vsini) by the true
rotational velocity. The latter quantity can be expressed as the
circumference divided by the rotation period ($v = 2\pi R_\star /         
\prot$). If the distance ($d$) to the star is known, the radius can be
expressed relative to the observable stellar angular diameter
($\Theta$) so that in terms of direct observables, \refeql{1} becomes
\begin{equation} \label{eq:2}
  \sin{i_\star} = \frac{v \sin{i_\star} P_{\rm rot}}{\pi \Theta d},
\end{equation}
where $\pi$\ refers to the mathematical constant (as opposed to the
parallax). The precision with which the sine of the stellar
inclination can be determined can thus be expressed in terms of the
uncertainties in these four observables:
\begin{equation} \label{eq:3}
  \bigg(\frac{\sigma_{\sin{i_\star}}}{\sin{i_\star}}\bigg)^{\!2} =
  \bigg(\frac{\sigma_{\vsini}}{\vsini}\bigg)^{\!2} +
  \bigg(\frac{\sigma_{\prot}}{\prot}\bigg)^{\!2} +
  \bigg(\frac{\sigma_{\Theta}}{\Theta}\bigg)^{\!2} +
  \bigg(\frac{\sigma_{d}}{d}\bigg)^{\!2}
\end{equation}
\noindent We note that \vsini\ and \prot\ will tend to be
inversely correlated, so there will be a contribution to the error 
from the covariance term, but this will be small compared to other 
error terms. Similarly, while the derivation of angular diameters is
potentially dependent on distance (more specifically, affected by
distance-dependent interstellar reddening), we argue that this
covariance can be accounted for through independent means, so we do
not consider a correlated error term. We discuss these issues in
more detail in \refsecl{errors}.

We leave the results in this form rather than taking the inverse sine
to obtain the inclination itself, because measurement errors will, in
some cases, result in an unphysical measurement of $\sini > 1$, which
clearly presents problems for estimating $i_\star$. If one must report
the inclination---e.g., for an individual system---this can be
addressed by working with the posteriors for $\sini$\ and $i_\star$ to
arrive at an appropriate confidence interval \citep[see][particularly
  the discussion of their Figure 2]{morton:2014}. For our purposes,
the expected distribution of $\sini$\ for various underlying obliquity
distributions can be modeled as easily as that of $i_\star$, and
comparisons can be made directly to the results obtained using
\refeql{2}, even though some fraction of systems will appear to have
$\sini > 1$.

In this paper, we consider the application of IPR for stars that TESS
is likely to survey for transiting planets. We select these stars
because TESS is the first of the large surveys proposed, and both
candidate input catalogs \citep{stassun:2014} and simulated stellar
and planetary populations \citep[][hereafter \sull]{sullivan:2015} exist
for this mission. A similar analysis could be conducted for PLATO 2.0
stars once its stellar sample is better defined. We also note that
because of the excellent astrometric precision of Gaia, this technique
will be applicable to the stars around which {\em Kepler} has already
found planets, despite their significantly greater typical
distances. Measurement precisions will determine the utility of IPR
for individual systems, whereas the power of a population analysis
will depend critically on obtaining a large sample of
measurements. With this in mind, we simulate results for two
observationally distinct samples:

\begin{enumerate}
    \item Stars with precisely measured angular diameters---the number
      of which will be set by the number of stars resolvable
      by current (and near-term) facilities, and is likely to be
      relatively small ($N \mysim 10^3$); and
    \item Stars with angular diameters estimated from photometric
      energy distributions or color relations---the number of which
      will be much greater ($N \mysim 10^5$) and, despite their lower
      precision, should be more amenable to a population analysis
\end{enumerate}
The results for Case (ii) will also depend upon the number of stars TESS
surveys. The simulated catalog of \sull\ contains $200,\!000$\ stars
monitored with $2$ minute cadence, and the authors note that
a large number of planets will be found using the full-frame images
(FFIs), which will sample $95\%$\ of the sky with a $30$ minute
cadence (similar to that of \kep). We directly consider only the short
cadence sample, but discuss the potential value of the larger sample from
the FFIs in \refsec{summary}.

Through Case (i), we hope to determine for which stars---and how many
stars---this method can provide a well constrained measure of the
obliquity, and through Case (ii), we hope to understand how well the
underlying obliquity distribution of small planets can be constrained
via such measurements.

\section{The Stellar Sample}
\label{sec:sample}

\subsection{Simulating TESS Targets}

\begin{figure}[!t]
\includegraphics[width=0.48\textwidth]{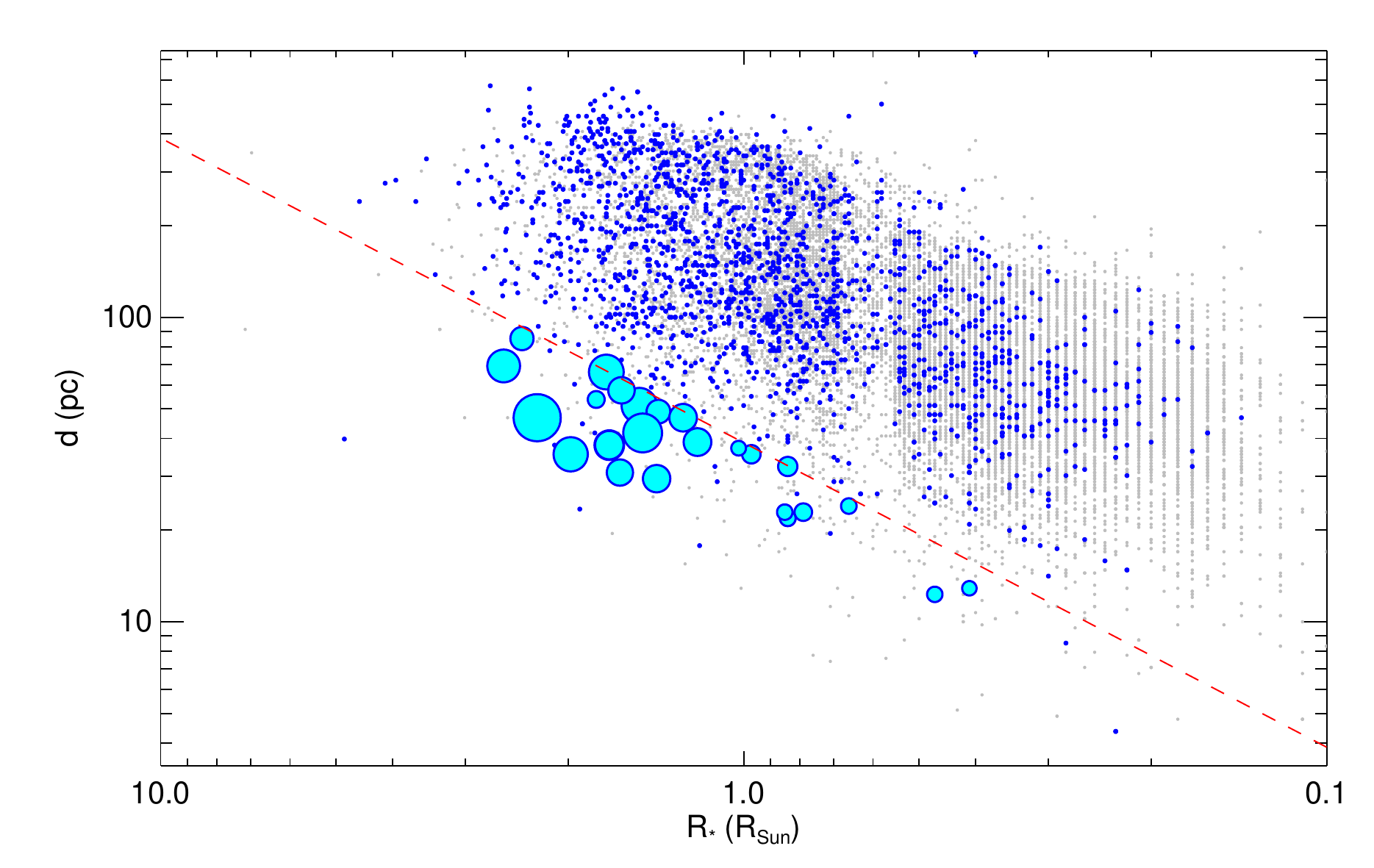}\hfill
\caption{ Distance as a function of stellar radius for the sample of
planet hosts. All $10$\ realizations are plotted here (gray dots). The
CHARA R-band resolution limit ($0.24$\ mas) is indicated with a red
dashed line. Stars that are resolvable lie below this line. Blue
dots indicate stars with measured \sini---i.e., those that pass our
observational cuts in \vsini\ and \prot\ and have detectable
photometric rotational modulation, as described in
\refsecl{n_meas}. Large cyan circles represent those stars that also
pass the resolution, magnitude ($R<10$), and declination ($\delta > 
-30\dg$) limits of the CHARA array---i.e., the subset with measured
\sini\ from directly measured diameters (Case (i)). The sizes of the
circles are scaled logarithmically by their projected rotational
velocities, which range from $2$ to $\mysim 120$\ \kms.
\label{fig:chara_cut}}
\end{figure}

\begin{figure*}[!t]
\includegraphics[width=0.975\textwidth]{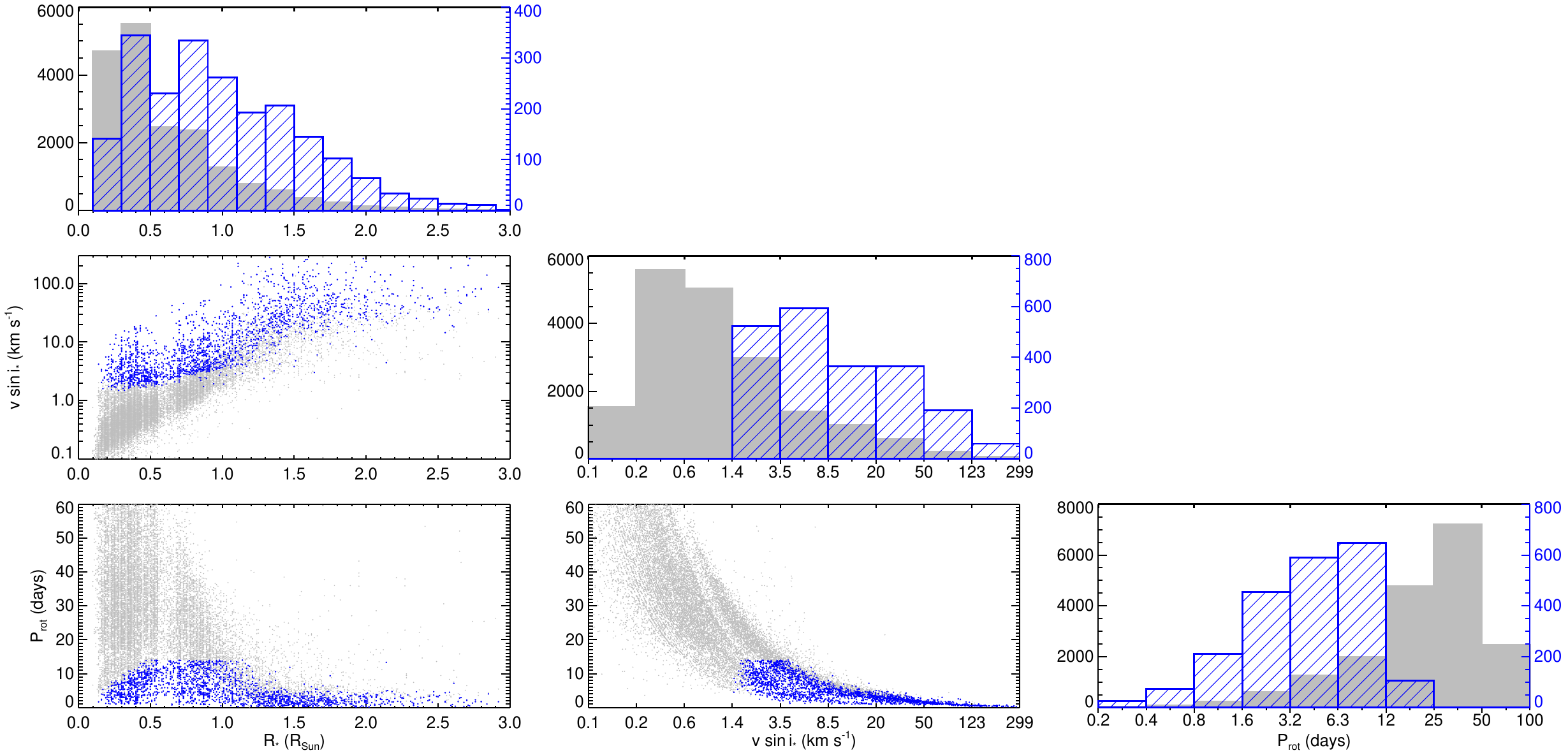}\hfill
\caption{ Comparison between the full sample of TESS planet hosts
  (gray) and the sample for which we expect \sini\ measurements
  (blue). We show correlation plots and histograms for \rstar, \vsini,
  and \prot. The right (blue) axes in the histograms correspond to the
  measured \sini\ sample, plotted as blue hatched histograms. All
  $10$\ realizations of the planet simulations are plotted, and we
  have assumed $f_{\rm al}=0.7$ to generate the plotted
  \vsini\ sample. It is clear that the smallest stars (slowest
  rotators) are most stongly affected by our observational cuts of
  $\vsini > 2$\ and $\prot < 14$.
\label{fig:sample}}
\end{figure*}

At the time of this publication, specific TESS targets have not yet
been selected, but there are working catalogs
\citep[e.g.,][]{stassun:2014}, from which at least $200,\!000$ dwarf stars will ultimately be selected, and a simulated
target sample and expected planetary population described by \sull. We
adopt the careful simulations of \sull, which employ the TRILEGAL
model of the galaxy \citep{girardi:2005} to generate the stellar
population, with slight modifications (e.g., to improve the agreement
of the simulated stellar radii with observations). Using this input
stellar catalog, the known frequency of exoplanets, and the expected
instrumental and survey characteristics of TESS, and a target
selection process based on planet detectability criteria, \sull\ then
simulate the stars around which TESS will detect planets and the types
of planets it will detect; they report one such planet catalog in
their Table 6. We used $10$\ realizations of the simulated TESS planet
catalog provided to us by P.  Sullivan to run the obliquity
simulations described below. One realization contains an average of
$1860$\ simulated planet hosts, and almost all are main sequence
stars. While the stars span spectral types B6 to M9, there are
relatively few early type hosts---$86\%$\ are later-type than G0
($\teff < 5920$\,K), and $57\%$\ are M dwarfs ($\teff < 3850$\,K). 
\reffigl{chara_cut} and \reffigl{sample} illustrate some
relevant properties of the planet hosts, but for a more detailed
description of the stellar and planetary samples, please refer to
\sull.

We also point out that while \sull\ treated stellar multiplicity with
care to ensure realistic TESS planet yields, we do not explicitly
include the effects of multiplicity in our calculations in the
following sections. The primary effect is likely to be a small bias in
some systems when fitting the photometric energy distribution to
estimate radii, but we adopt conservative errors to mitigate this.

\subsection{Calculating the Observable Quantities}

The stellar properties provided by \sull\ include the distance
modulus, \rstar, \teff, and absolute $V$ magnitude, which we can use
to derive the observable quantities from \refeql{2}. We first
trivially calculate $d$\ from the distance modulus and $\Theta$\ from
\rstar\ and $d$.

Next, we use the gyrochronological relation from \citet{mamajek:2008}
to calculate rotation periods:
\begin{equation} \label{eq:4}
        \prot = 0.407\ [(B-V) - 0.495]^{0.325}\ t^{0.566}
\end{equation}
For this, we need the $B$ magnitudes, which we derive using the
\teff--color table presented by E.
Mamajek\footnote{\url{http://www.pas.rochester.edu/~emamajek/EEM_dwarf_UBVIJHK_colors_Teff.txt}}
and \citet{pecaut:2013}. There is significant uncertainty in the four
constants in \refeql{4} (because real stars display a scatter about
this relation), so we draw a new set of the four constants for each
star, using the errors on these constants reported by the
authors. This produces a stellar sample with a realistic scatter of
rotation periods about the gyrochronological relation. We further note
that the sample includes a handful of early type stars and a large
number of M dwarfs, but the above relation is only calibrated for $0.5 
\lsim (B-V) \lsim 0.9$\ (F7-K2). \citet{mamajek:2008} note that
F3V--F6V spectral types appear to be where gyrochronology fails due to
the thinning of convective envelopes and the loss of magnetic dynamos,
thus preventing the magnetic braking from which spin-down laws
emerge. These earlier types should therefore spin faster than their
slightly less massive counterparts. As an approximation, we assign
rotation periods for earlier types by assuming $(B-V)=0.5$, which may
slightly overestimate the rotation periods for the hottest stars. For
late-type dwarfs with $(B-V) > 0.9$, we simply apply \refeql{4} and
acknowledge that these less massive, and in some cases fully
convective, stars probably obey a somewhat different spin-down law. We
argue that this is unlikely to significantly bias our
results. Qualitatively, cooler stars will spin more slowly, so the
declining rotation period relation should be correct to first
order. Moreover, some very late-type stars are expected to have
significant rotation---the predictable, monotonic spin-down seen in
solar-type stars gives way to a bimodality for later types that
includes both fast and slow rotators
\citep[e.g.,][]{irwin:2011,mcquillan:2014}. While this bimodality is
most pronounced for the coolest stars, it begins to emerge even for
stars as hot as $\mysim 4500$\,K. At even lower masses, many brown
dwarfs rotate with periods on the order of hours \citep[e.g.,][and
  references therein]{metchev:2015}. In all cases, by erring toward
overestimated rotation periods, we will also overestimate
uncertainties for \prot\ and \vsini, thereby obtaining more
conservative results. That said, one must also worry that comparisons
between observed data and simulated data could introduce
systematically biased results if the properties of the simulated data
do not reflect the properties of real stars. Since \refeql{4} is
expected to break down for late-type stars, we explore simulations of
a more conservative sample selection --- one for which gyrochronology
is better suited --- in \refsecl{conservative}.

A further complication in simulating \prot\ is that measured rotation
periods (including those used to derive gyrochronology relations)
track the rotational velocity at the active spot latitudes rather than
at the equator---the rotation periods we assigned are therefore the
periods at the active latitudes. We can assign {\em equatorial}
rotation periods by assuming a differential rotation law and spot
latitude distribution. Following the approach of \citet{hirano:2014}
and \citet{morton:2014}, we adopt the functional form for differential
rotation from \citet{cameron:2007}:
\begin{equation} \label{eq:5}
        \prot(\ell) = \frac{P_{\rm rot,eq}}{1-\alpha\sin^2\ell},
\end{equation}
and we draw spot latitudes from a distribution $\ell = 20 \pm 
20\dg$, and assume the strength of differential rotation is similar
to the Sun ($\alpha=0.23$).

\citet{hirano:2014} showed that differential rotation will similarly
result in a measured \vsini\ that is too large by a factor $\mysim(1 - 
\alpha/2)^{-1}$. While applying this correction to observations of
real systems has the potential to introduce a small error if the true
strength of differential rotation is significantly different from the
assumed solar value of $\alpha=0.23$, in our simulations the true and
observed values would both receive the same correction, so we ignore
this small factor.

With the equatorial \prot\ and \rstar\ in hand, we can generate
\vsini\ by drawing $i_\star$\ from an appropriate distribution. We
note that an isotropic distribution is most likely not the correct
choice for this distribution---our hypothetical planets all transit,
so if $\psi$\ is preferentially low (well-aligned), then $i_\star$\
must be preferentially high (edge-on rotation). We explore the effect
of different obliquity distributions in \refsecl{results}.

\section{Assigning Uncertainties and Simulating Observations}
\label{sec:errors}

\subsection{Distance Uncertainty}
\label{sec:distance}

The most uniform and complete set of distances currently available for
nearby Sun-like stars (${<}100$\,pc) comes from the Hipparcos mission
\citep{esa:1997}. \citet{leeuwen:2007} quoted a mean parallax
precision for single stars of $0.33$\ mas for $H_P < 7$\ mag, and
$0.56$\ mas for $H_P < 9$\ mag, which corresponds to distance
precisions of $\mysim 1\%$\ for Sun-like stars at $30$\ pc, and
$\mysim 4\%$\ for those at $75$\ pc. For many of the fainter TESS
targets, this uncertainty easily becomes the dominant term in the
error budget of \refeql{3}. Fortunately, the Gaia mission
\citep{perryman:2001} promises to improve upon this considerably, by
providing parallax measurements to roughly 1 billion stars; first
estimates of these distances are expected to be available prior to the
launch of TESS. \citet{bruijne:2012} quoted an anticipated precision
of better than $10\ \mu$as for $G<12$\ mag stars--- corresponding to a
distance precision better than $0.1\%$\ at $100$\ pc and better than
$1\%$\ at $1$\ kpc---although faint stars and some very close visual
binaries may carry additional measurement errors. Evaluation of
in-orbit commissioning data suggests that despite some minor
operational hiccups (e.g., significant stray light levels, larger than
expected instability of the basic angle between the lines of
sight of the two telescopes) the astrometric performance will remain
within the $20\%$\ science margin set before launch; the updated
performance predictions given by \citet{bruijne:2014}\footnote{See
  also a summary on the mission website:
  \url{http://www.cosmos.esa.int/web/gaia/science-performance}.}
suggest that the majority of the TESS stars will carry a parallax
error smaller than $\mysim 15\ \mu$as. We approximate distance errors
by propagating a constant $15\ \mu$as uncertainty in parallax,
$\sigma_d = 15\times 10^{-6} d^2$\ pc.

\subsection{Stellar Angular Diameters: Available Populations and Uncertainties}
\label{sec:diameter}

Stellar angular diameters can either be measured directly from
interferometric measurements or estimated from bolometric fluxes and
effective temperatures: $\Theta^2 = 4 F_{\rm bol} / (\sigma T_{\rm 
eff}^4$). Since the directly measured values translate to stellar
inclinations with the smallest uncertainties, we consider this
population first. For a star to be spatially resolved, it must be
both bright enough for precise measurements and larger than the
resolution limit. Currently the world's longest baseline
interferometer operating at optical and infrared wavelengths, where
photospheres are the brightest, is Georgia State University's CHARA
Array, located on Mt. Wilson in California \citep{brummelaar:2005}.
With a long baseline of $331$\ m, the Array yields an angular
resolution ($\lambda / 2B$) of $0.51$\ mas in the H band and
$0.24$\ mas in the R band; an adaptive optics system is currently
being installed that is predicted to achieve limiting magnitudes of
$9.0$\ at H and $10.0$\ at R \citep{brummelaar:2012, brummelaar:2014};
current limiting magnitudes for precision measurements are $7.0$\ and
$7.0$, respectively.

To determine the number of stars that the CHARA Array is likely to
spatially resolve, we adopt the R band adaptive optics limits of
$\theta_{\rm res} = 0.24$\ mas and $R_{\rm lim} = 10.0$\ mag, and
assume a declination limit of $\delta > -30\dg$. These limits are
applied to the $10$\ realizations of the TESS simulated planetary
population, and yield a mean of $14.8$\ TESS transiting planet hosts
resolvable by CHARA. This subset of stars is illustrated in
\reffigl{chara_cut}. We note that for all but the very latest types,
the primary limitation is their small angular sizes, rather than their
faint $R$ magnitudes; relative to the CHARA limits, the typical TESS
planet host is small but not necessarily faint. The brightness of
these stars will enable high signal-to-noise ratio observations that are important for
achieving precise size measurements, and their small sizes are not an
insurmountable obstacle; \citet{huber:2012}, for example, obtained
angular diameters to within $2\%$\ for several bright stars close to
(and even beyond) the nominal resolution limit using a modest number
of observations. It thus seems plausible that the majority of these
stars that are large enough and bright enough for CHARA to resolve and
could have their sizes measured with similar precisions. We therefore
adopt a $2\%$\ uncertainty in directly measured angular diameters.

Since the subsample that can be spatially resolved with CHARA
represents only a small fraction of the expected TESS planet hosts, we
also consider the precision with which radii can be estimated.
\citet{boyajian:2014} demonstrate that angular diameters can be
predicted with a precision of $5\%$\ from single color relations and
an absolute magnitude, based on the dispersions about the best fit
relation for nearby stars with interferometrically measured radii and
Hipparcos distances. However, applying these relations blindly to TESS
targets will likely yield biased results since the sample extends to
distances of $\mysim 500$\,pc (see \reffigl{chara_cut}) and many stars
likely experience modest extinction. There are two promising methods
to account and correct for this. First, the precise distances that
Gaia will provide, coupled with spectroscopically determined
temperatures, will yield all sky extinction maps
\citep{schultheis:2015} and enable first order corrections. And for
planet hosts with a spectroscopically determined temperature, their
observed photometric energy distribution can be compared to that
expected for their temperature to derive their line-of-sight
extinction. Thus, with very accurate Gaia distances and extinction
corrections, photometric energy distributions (as opposed to single
color relations) can be used to get accurate bolometric fluxes. When
combined with spectroscopically determined temperatures, these will
yield accurate angular diameters via the Stefan--Boltzmann
Law---$\Theta \propto \left(F_{\rm bol}\right)^{0.5} 
\left(\teff\right)^{-2}$. We adopt a conservative $5\%$\ uncertainty
in estimated angular diameters, since that precision has already been
demonstrated using single colors, but we expect that it should be
possible to do better using the full photometric energy distribution.

\subsection{Rotation Period Uncertainty}
\label{sec:period}

The photometric precision and high cadence ($\mysim 
2$\ minute) observations of TESS will enable the detection of a large number of
stellar rotation periods. For example, \citet{mcquillan:2014} report
$34,\!000$ rotation periods among the \kep\ sample, corresponding to a
recovery rate of $\mysim 25\%$ for cool (${<}6500$\,K) dwarfs without
eclipsing binary companions.

The TESS and \kep\ dwarf samples will differ in several ways, and
while we discuss these differences below, we ultimately argue that
adopting detection rates from the \kep\ sample is justified in our
simulations. Perhaps the most important difference between data sets
will be the typical time span of observations for TESS stars. In
general, they will not have the luxury of the long-term monitoring
that \kep\ provided---most TESS stars will be monitored for only
$27.4$\ days (one spacecraft pointing), though near the ecliptic
poles, field overlap will allow $\mysim 1$~year of continuous
monitoring. We therefore expect that precise rotation periods will
only be determined for $\prot < 14$\ days, and remove slower rotators
from our simulations. This observational constraint is likely to limit
TESS detection of rotation periods mainly to stars younger or
earlier-type than the Sun (i.e., rotating faster than $P_{\rm rot, 
\odot} \mysim 26$\ days), and may hinder detections for
slowly rotating late-type stars. However, in the context of
\kep\ rotation periods, the former limitation may not be important, as
\citet{mcquillan:2014} suggest that very few rotation periods of stars
older than the Sun were detected. Similarly, in the context of Case (i)
(interferometrically determined $\Theta$), relatively few late type
stars will be resolvable because of their small sizes and large
magnitudes; reduced sensitivity to rotation periods of late-type stars
will have relatively little effect on the overall results. For Case (ii)
(photometrically determined $\Theta$), TESS might not observe multiple
full rotations for old or cool stars, which could hinder the
application of IPR. However, the typical amplitudes of variation are
large enough ($\mysim 1$\ to $22$\ mmag in the $5^{\rm th}$\ to
$95^{\rm th}$\ percentile) that ground-based follow-up should in most
cases be capable of determining the rotation periods. This would not
be ideal, but is not an insurmountable obstacle.

Because TESS will observe the brightest stars in the sky, it is worth
comparing TESS and \kep\ photon noise. TESS targets will typically be
$30$--$100$\ times brighter than \kep\ targets. However, the TESS
aperture has an area $100$\ times smaller than \kep\ and photon noise
will therefore generally be comparable to or greater than for
\kep. \kep's longer duration is also an asset to beat down noise, and
the results will ultimately depend on the photometric precision TESS
is able to achieve. The brightness of TESS targets will at least
better enable ground-based follow-up, e.g., to determine or improve
rotation periods of stars for which TESS does not observe many
rotation periods.

We assume that the fractional uncertainties in the measured
\prot\ reported by \citet{mcquillan:2014} are also roughly indicative
of the precisions one can expect from TESS, at least for periods short
enough to be determined reliably. This assumption may ultimately prove
to be optimistic, so we also run simulations that adopt a more
conservative \prot\ error estimate in \refsecl{conservative}. Here,
however, we assume that the combination of sophisticated analysis and
photometric follow-up can provide rotation periods with precisions
similar to those determined using \kep\ data. Among $34,\!000$\ main
sequence \kep\ stars for which \citet{mcquillan:2014} detect rotation
periods, the highest probability density (and $1 \sigma$\ error) of
the fractional period uncertainty is $0.29\%^{+1.51\%}_{-0.29\%}$; the
period uncertainty is on average dependent upon the rotation period
itself. To account for this, we assign uncertainties as a function of
rotation period using the \citet{mcquillan:2014} catalog. For each bin
in \prot, we calculate the probability density function (PDF) of the
fractional uncertainty using a kernel density estimator. For each star
in that bin, we then draw a fractional uncertainty according to the
PDF, which is trivially converted to an absolute uncertainty,
$\sigma_\prot$.  One complication is that the observed rotation period
is actually the period at the active spot latitude rather than
equatorial rotation.  To account for this we first simulate the
observed rotation period, $\hat{P}_{\rm rot}(\ell) = P_{\rm rot}(\ell) 
+ x\sigma_\prot$, and then assume $\ell = 20\dg$\ in order to
estimate the correction needed to determine the equatorial rotation
period: $\hat{P}_{\rm rot} = \hat{P}_{\rm rot}(\ell) (1 - 
\alpha\sin^2(20\ {\rm deg})) = 0.973 \hat{P}_{\rm rot}(\ell)$. The
cumulative effect of having to estimate the active spot latitude
correction is typically an additional error in the equatorial rotation
period of $\mysim 2.5\%$\ (stemming from the uncertainty in $\ell$)
that is not included explicitly in the measurement errors for the
observed rotation period.

\subsection{Projected Rotational Velocity Uncertainty}
\label{sec:vsini}

The projected (equatorial) rotational velocities, \vsini, can be
obtained from high resolution spectroscopy. Unlike \teff, \logg, and
\feh, \vsini\ is mostly free from degeneracy with the other
parameters. For stars rotating sufficiently fast such that rotation is
the dominant broadening mechanism (\vsini $\gtrsim$ 10), \vsini\ can
typically be determined to a few percent with high signal-to-noise
ratio, high dispersion spectra
\citep[e.g.,][]{siverd:2012,bieryla:2015}. However, for more slowly
rotating stars, complicating effects such as the assumed
macroturbulent velocities of convective cells (\vmac) can directly
affect the predicted spectral line profiles. For example, the
difference in derived \vsini\ if one assumes \vmac\ values of $2$\ and
$3$\ \kms\ is $\mysim 0.25$\ \kms, or $2.5\%$ for a star with
$10$\ \kms\ total line broadening. There is hope that observationally
calibrated relationships between \teff, \logg, and \vmac\ \citep[e.g.,
  the asteroseismically calibrated relationship presented
  by][]{doyle:2014} can be used to constrain \vmac\ and mitigate such
a bias. In practice, however, even with perfect knowledge of \vmac,
current typical measurement uncertainties for stars with $\vsini 
<10$\ \kms\ are $0.2-0.3$\ \kms, and thus approach a
$10\%$\ uncertainty for stars with \vsini\ values of only a few
\kms\ \citep[e.g.][]{gray:2013, tsantaki:2014}. We note that some more
sophisticated analysis techniques may improve the precision several
fold \citep{gray:1997} if measuring \vsini, as opposed to simply
accounting for it in a stellar model, were the goal. Finally,
measurement of \vsini\ values that are less than the instrumental
resolution are possible, but become even more challenging. The HARPS-N
spectrograph \citep{cosentino:2012}, for example, has a resolving
power of $R \mysim 115,\!000$, corresponding to a resolution element
of $2.6$\ \kms; if \vsini\ is much slower than this, then it is only a
minor contributor to the total line broadening. Large fractional
changes in \vsini\ therefore have relatively little effect on the line
profiles and cannot be measured precisely. In these cases it is best
to assign an upper limit on the \vsini\ value, although a proper
choice for this limit, relative to the instrumental resolution, is
also data quality dependent.

Given the above considerations, we adopt the following uncertainties
for \vsini. For stars with $\vsini \ge 10\ \kms$, we assume an
uncertainty of $2.5\%$. For stars with $10\ \kms \ge \vsini \ge 
2\ \kms$, we assume a constant uncertainty of $0.25$\ \kms, which
corresponds to an uncertainty of $2.5\%$\ at $10$\ \kms\ and
$12.5\%$\ at $2$\ \kms. For stars with $\vsini < 2$\ \kms, we assume
that only an upper limit can be inferred, and we remove these stars
from our sample.

We acknowledge that much of the exoplanet literature quotes larger
errors for slow rotators (often $\pm 0.5$\ \kms), but we point out
that this convention has arisen because the effort and/or telescope
resources required to push toward more precise \vsini\ would usually
provide diminishing scientific returns; in most cases, improving the
\vsini\ precision would not affect the conclusions of the paper at
all, so a adopting a conservative error makes sense. Achieving our
assumed uncertainties may be a difficult pursuit, but it is one that
is feasible with current techniques and in the context of this work,
the scientific returns would be well worth the effort. Finally, we
note that a concerted effort to make these measurements would not have
to be carried out for all TESS hosts, but only the much smaller number
to which IPR can be applied.

\subsection{Simulating Measured Quantities}
\label{sec:sim_obs}

To determine which, and how many, systems will be amenable to
projected obliquity measurements via IPR, we simulate observations of
each star in our sample to arrive at simulated observed values of
$\sini$. This simulated observed distribution has two uses. First, it
can be populated with transiting planets according to realistic
occurrence rates in order to calculate the likely yield of projected
obliquities (and their distribution and uncertainties). Second, by
running this simulation many times, we can build well-sampled,
expected distributions of observed $\sini$\ values under various
assumptions (number and characteristics of survey targets,
observational uncertainties and limits, etc.). These distributions can
be used as comparison samples against which we can compare our
observed distributions (e.g., in a Kolmogorov--Smirnov test). This
will therefore allow us to test observed distributions against
arbitrary underlying obliquity distributions.

We first point out that measurement precision as a function of stellar
properties is somewhat nuanced, as the precision of several of the
observables is expected to have a dependence on other stellar
properties. We therefore must be careful when simulating observed
quantities and their measurement uncertainties. Very generally, given
the observable quantities $p=\{d,\Theta,\prot,\vsini\}$\ with true
values $p_i$ as calculated in \refsecl{sample}, we need to assign for
each star in our sample the observed values, $\hat{p}_i$, and
uncertainties, $\sigma_{p_i}$. Under the assumption of uncorrelated
Gaussian errors, these quantities can be related simply by:
\begin{equation} \label{eq:6}
        \hat{p}_i = p_i + x \sigma_{p_i},
\end{equation}
where $x$\ is drawn from a normal distribution,
$x\mysim\mathcal{N}(0,1)$.

Using the above relation we can simulate measured quantities for $d$, 
$\Theta$, and $\prot$. Simulating measurements of \vsini, however,
requires assigning a distribution of $i_\star$ values, which are
unlikely to be random for stars that host transiting planets. For
example, if most planets are coplanar with the stellar equator, then
for a sample of transiting planets, the underlying inclination
distribution should be confined to near-$90\dg$\ 
orientations. However, if all systems experience significant evolution
of their angular momenta, an isotropic distribution might be more
appropriate. The true distribution may contain a mix of systems:
well-aligned systems that have formed in an equatorial disk and
experienced calm migration (or no migration at all), plus a more
isotropic distribution caused by interaction of the planets or
proto-planetary disk with a binary or additional planetary
companions. For Case (i) (interferometric $\Theta$), we do not expect a
large number of systems (or a powerful population analysis), and we
thus confine our investigation to two simple inclination
distributions: isotropic, in which the probability of observing a
particular inclination is given by $P(i_\star)\mysim\cos{i_\star}$;
and well-aligned, which we characterize as an isotropic distribution
truncated to within $10\dg$\ of alignment. For Case (ii) (photometric
$\Theta$), we expect many more systems, and therefore anticipate that
we will be able to distinguish much more finely between underlying
distributions. We therefore simulate many "mixed" distributions,
comprised of varying fractions of the above isotropic and aligned
distributions. We refer to these mixed distributions by the fraction
of aligned systems---for a mixed distribution with $f_{\rm al} = 0.7$,
$70\%$\ of systems are drawn from an aligned distribution and $30\%$\
of systems are drawn from an isotropic distribution. Regardless of the
distribution from which $i_\star$\ is drawn, our observed projected
rotational velocities are simply given by $\hat{v}\sini = \vsini + 
x\sigma_\vsini$\ (with $\sigma_\vsini$\ chosen as outlined in
\refsecl{vsini}).

For a given distribution of $i_\star$ values, all four measured
quantities can be simulated and the uncertainty in $\sin{i_\star}$ can
be determined from formal error propagation.

\subsection{Resulting Uncertainty on the Stellar Inclination}
\label{sec:sigma}

As described in more detail above, a typical spatially resolvable star
might have errors of $\sigma_\prot \mysim 1\%$, $\sigma_\theta \mysim 
2\%$, and $\sigma_d \mysim 0.1\%$, with an additional $\mysim 
2.5\%$\ uncertainty in $\sigma_\prot$\ introduced by correcting for
differential rotation. The fractional error on \vsini\ depends on
\vsini\ itself; for the example of a star rotating at $5 \pm 
0.25$\ \kms\ ($\sigma_\vsini \mysim 5\%$), this combination of
uncertainties leads to $\sigma_\sini \mysim 6\%$. The corresponding
constraint on $i_\star$\ would depend on the inclination: for nearly
edge-on rotation ($\sin{i_\star} \mysim 1$), one would derive an
uncertainty in $i_\star$\ of $\mysim 20\dg$, but for lower stellar
inclinations, the uncertainty may be only a few degrees. Thus,
individual well-aligned systems will still be difficult to identify
with confidence---especially around slow rotators for which the
\vsini\ fractional uncertainty is large---but misaligned systems
should result in highly significant detections and precise
measurements. Despite this difficulty for some individual systems, the
{\em population} of $\sini$\ measurements should be able to constrain
the underlying obliquity distribution, and the smaller uncertainties
resulting from IPR using precise distances and angular diameters
should reduce the number of measurements required to do so.

\section{Results}
\label{sec:results}

\subsection{Expected Number of Obliquity Measurements}
\label{sec:n_meas}

The above discussion quantifies how well the stellar inclination can
be determined in any one case, and the simulated stellar population
allows us to estimate the number of transiting systems for which these
projected obliquity measurements can be realized. This number can be
estimated by multiplying the initial number of targets successively by
the fraction that hosts planets, the fraction for which a transit is
detected, and the fraction that has detectable rotational modulation
and rotational line broadening:
\begin{equation} \label{eq:7}
        N_{\rm ob} = N_\star \times f_{\rm pl} \times p_{\rm tr} \times f_{\rm rot}.
\end{equation}
By using their simulated TESS planet catalog, we have implicitly
adopted the first three terms from \sull; briefly, they assume
$N_\star=200,\!000$ and assign planet occurrence rates ($f_{\rm pl} 
\times p_{\rm tr}$) as a function of stellar properties and orbital
period using the \kep\ planet occurrence statistics reported by
\citet{fressin:2013} and \citet{dressing:2013,dressing:2015}. Given
the mean number of planets from $10$\ realizations of the
\sull\ simulations, \refeql{7} becomes $N_{\rm ob} = 1863 \times 
f_{\rm rot}$.

Not all stars are active enough to reveal their rotation periods, and
activity level varies as a function of spectral type, so we cannot
adopt a single value for $f_{\rm rot}$ in our
simulations. Additionally, our fiducial simulation makes the
assumption that \vsini\ cannot be measured below $2$\ \kms\ and that
rotation periods longer than $14$\ days will not be identified by
TESS, so systems that do not satisfy these criteria will be removed,
even if the photometric amplitude of rotational modulation is expected
to be detectable. To assess detectability of rotational modulation, we
first assume that TESS will recover rotation periods at the same rate
as \kep. \citet{mcquillan:2014} detected rotation in $83\%$\ of cool
stars (${<}4000$\,K), but in only $\mysim 20\%$\ of stars hotter than
the Sun. To enforce this condition, we divide our sample into bins of
$100$\ K in \teff\ and then flag stars as detectable with probability
according to the PDF of the \citet{mcquillan:2014} rotation periods in
that \teff\ bin, normalized to the detection rate for that \teff. In
this way, our final distributions of rotation periods and detection
rates as a function of \teff\ broadly match the \kep\ sample before we
apply our observational cuts of $\prot<14$\ days and
$\vsini>2$\ \kms. Finally, we note that the fraction of transiting
planet host stars with $\vsini > 2$\ \kms\ depends upon the underlying
obliquity distribution: well-aligned systems will have larger
\vsini\ on average. This effect is seen in the results of our
simulations. After assessing \prot\ detectability and applying the
observational cuts, we estimate $N_{\rm ob} \mysim 189$\ for an
isotropic distribution of obliquities and $N_{\rm ob} \mysim 214$\ for
a well-aligned distribution.

We compare the properties of the full planet host sample to those of
the ``measured \sini'' sample in \reffigl{sample}. One can clearly see
the primary result of the observational limits we have imposed: the
excluded slow rotators tend to be the smallest stars, so the
``measured \sini'' sample is biased toward larger (more massive) stars
than the full planet host sample. Furthermore, the M dwarfs for which
\sini\ {\em will} be measured belong to the rapidly rotating subset of
M dwarfs, which may indicate comparative youth or perhaps more
fundamental differences. The properties of planetary systems orbiting
these stars may therefore not be representative of all M dwarf
planets. This will be important to keep in mind when interpreting
results. This also highlights an opportunity -- the extension of
observational limits to smaller \vsini\ and larger \prot\ through
improved techniques and follow-up observations will provide a larger
{\em and more representative} population for this analysis. We discuss
this opportunity further in \refsecl{apps}.

\subsection{Constraints on the Underlying Obliquity Distribution from Interferometric Measurements}
\label{sec:constraints_chara}

The discussion in \refsecl{n_meas} considers the large sample of Case
(ii) (photometrically derived angular diameters), but the (small) number
of CHARA-resolvable TESS planet hosts can also be determined directly
from the distances and radii provided by the \sull\ simulations. A mean
of $14.8$\ resolvable planet hosts are expected, but only $2.2$\
(isotropic distribution) or $2.5$\ (aligned distribution) are expected
to have $\vsini > 2$\ \kms\ and a detected $\prot < 14$\ days.

A handful of projected obliquity measurements is not sufficient for a
population analysis, but this group of planetary systems is likely to
be well-studied, since uncertainties in stellar properties are most
often the strongest limitation on our understanding of the properties
of transiting planets, and the precision afforded by directly measured
radii allows greater insights into processes governing the structure
and evolution of both stars and planets
\citep[e.g.,][]{boyajian:2015}. It is therefore plausible that through
careful follow-up observations, another handful of these $\mysim 15$\
systems will have rotation periods and rotational velocities measured,
even beyond the $14$\ day and $2$\ \kms\ limits assumed above. In any
case, for a strong constraint on the underlying obliquity
distribution, we must turn to the larger sample of photometrically
derived angular diameters.

\subsection{Constraints on the Underlying Obliquity Distribution from Photometric Measurements}
\label{sec:constraints_phot}

Determining $\Theta$\ from the photometric energy distribution or
color relations will be harder to apply to individual systems because
of its less precise measurements, but it holds greater promise as a
means to probe the population as a whole because we expect hundreds of
systems to be amenable to this technique. We describe here one
methodology that can be used to characterize the true obliquity
distribution and demonstrate it using our simulated observations of
TESS planet hosts.

\begin{figure}[!t]
\includegraphics[width=0.48\textwidth]{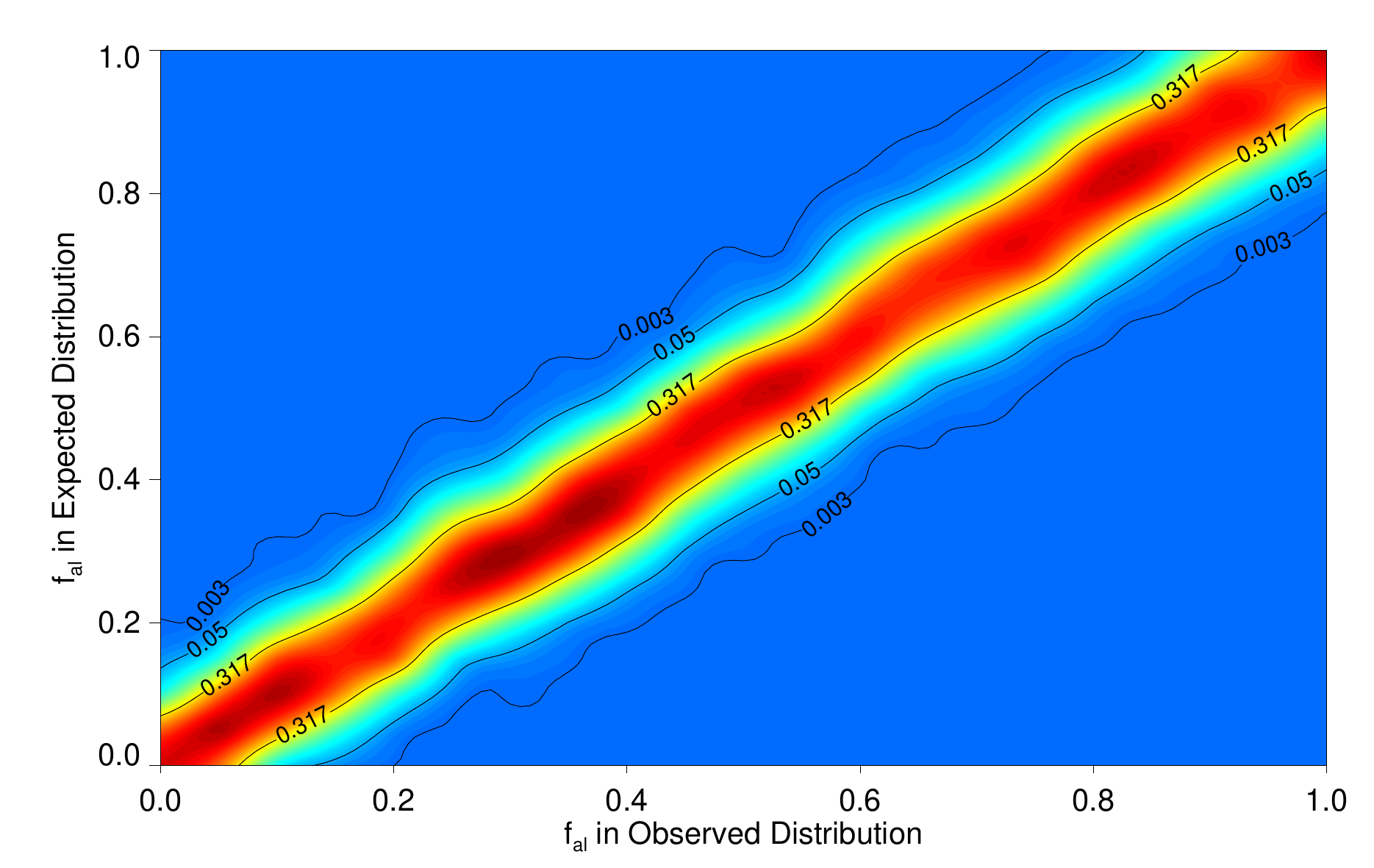}\hfill
\caption{ KS test $p$-values (colored contours) for observed and
  expected \sini\ distributions comprised of $f_{\rm al}$\
  well-aligned systems (and $1-f_{\rm al}$ randomly oriented
  systems). Low $p$-values (blue regions) correspond to pairs of
  expected and observed distributions that are unlikely to arise from
  the same parent distribution. The most likely true fraction of
  aligned systems is that for which the expected distribution produces
  the largest $p$-value (red regions), and the distance over which the
  $p$-values decline characterizes the precision with which the true
  aligned fraction can be measured. For each observed distribution, we
  adopt the region with $p>0.05$\ (red to cyan) as the uncertainty on
  $f_{\rm al}$\ for the true underlying distribution. The results are
  summarized numerically in \reftabl{ks_phot}.
\label{fig:ks_phot}}
\end{figure}

Given an underlying obliquity distribution, one may simulate a stellar
and planetary population and a set of observations of those systems in
order to construct the distribution of \sini\ values that one expects
to recover through observation. We refer to this as the ``expected 
distribution''. It is a distribution corresponding to the true
underlying distribution once it has been subjected to our
observational constraints (i.e., $\vsini > 2$\ \kms, $\prot < 
14$\ days) and measurement errors. This expected distribution can then
be used as a comparison against the \sini\ distribution that arises
from observations of TESS systems---the ``observed 
distribution''. (Since TESS has not yet launched, we also simulate the
observed distribution here, of course.) To constrain the true
underlying distribution, we run a series of Kolmogorov--Smirnov (KS)
tests to characterize the similarity between the observed distribution
(for a given value of $f_{\rm al}$) and many expected distributions
(for many different values of $f_{\rm al}$). For pairs with $p<0.05$\
($<5\%$\ chance that the same parent distribution would produce both
the observed and expected distributions), we consider the
distributions inconsistent. For a given observed distribution, we
report the best fit as the expected distribution with the largest
$p$-value and assign errors to encompass the expected distributions
that produced $p>0.05$. \reffigl{ks_phot} illustrates this comparison
for our fiducial simulations and a range of underlying obliquity
distributions. The results are summarized in \reftabl{ks_phot}. It can
be seen clearly that the best fit distribution does indeed accurately
recover the distribution from which our systems were drawn.

One advantage of the technique outlined herein is that by simulating
the stellar sample, the observations, and ultimately the expected
distribution, the observed distribution can be tested against any
model distribution, and the expected distribution captures the
observational biases that would not be accounted for in a direct fit
to an assumed functional form for the true distribution. In practice,
it makes sense to test physically motivated distributions---such as
our examples that have been parametrized to include $f_{\rm al}$\
aligned systems and $1-f_{\rm al}$\ isotropic systems, or the Fisher
distribution favored by \citet{morton:2014}---but the model
distributions need not be parametric, and the use of the KS test to
reject model distributions allows a fully non-parametric analysis. One
could, for example, directly use the results of arbitrarily complex
dynamical evolution models as the comparison distributions.

A potential drawback to this technique is that the reported errors
from the grid of KS tests are difficult to interpret in a standard
way. To investigate how realistic the errors are, we also fit the
observed cumulative distribution function (CDF) directly to a model
CDF. For nodes in the observed CDF located at $(\hat{x},\hat{y})$, the
corresponding nodes in the model CDF are located at $(\hat{x},y)$. The
model CDF is taken to be a linear combination of aligned and isotropic
CDFs:
\begin{equation} \label{eq:8}
\begin{split}
  {\bm y}(\hat{\bm x};f_{\rm al}) &= \frac{1}{\beta} \Big( \alpha_1 f_{\rm al}\ {\bm y}(\hat{\bm x};1) + \alpha_0 (1-f_{\rm al})\ {\bm y}(\hat{\bm x};0)\Big) ; \\
                  \beta &= \alpha_1 f_{\rm al} + \alpha_0 (1-f_{\rm al}),
\end{split}
\end{equation}
where the $\alpha$\ terms are weight parameters to account for more
obliquities being measured in aligned distributions, and $\beta$\ is
simply the normalization. We fit to minimize $\chi^2$, with errors for
the observed CDF at a given \sini\ value assigned to be proportional
to the square root of the \sini\ PDF (equivalent to counting
statistics), and scaled so that $\chi^2_{\rm red}=1$. The
$1$-$\sigma$\ errors from this direct fit, corresponding to the
central $68.3\%$\ highest probability density region, are typically
about $\pm 0.09$. These errors roughly correspond to the errors one
would get from the KS grid by assuming the $1$-$\sigma$\ region is
well-described by $p>0.317$. We therefore conclude that our reported
errors are, if anything, conservative. We choose to retain the larger
errors, as potential systematics in our method (discussed in
\refsecl{apps}) may reduce accuracy for real observations.

Based on these simulations, we conclude that the expected yield of
projected obliquity measurements from TESS will enable identification
of the likely true obliquity distribution for hosts of planets of all
sizes. In our example---motivated by migration processes that might be
expected to produce a bimodal distribution of obliquities (i.e.,
aligned or isotropic, depending on the migration mechanism or lack
thereof)---TESS obliquities should determine the fraction of systems
drawn from a well-aligned distribution to within $\mysim 0.15$. The
results of such an analysis holds insight into the relative importance
of various proposed formation and migration mechanisms.

\begin{deluxetable}{lr}
  \tablewidth{\textwidth}
  \tablecaption{\\Constraints on the Underlying Obliquity Distribution\vspace{-0.5cm}
    \label{tab:ks_phot}
  }
  \tablehead{
  	\multicolumn{2}{c}{\vspace{-0.6cm}} \\
 	\colhead{{\kern -1.1cm} $f_{\rm al}$ drawn{\kern 0.1cm}} &
    \colhead{{\kern 0.1cm}$f_{\rm al}$ derived{\kern -1.1cm}}
  }
  \startdata
  ${\kern -1cm}  0\%$ & $  0^{+14}_{-0}\% {\kern -1cm}$ \\
  ${\kern -1cm} 10\%$ & $ 10^{+15}_{-10}\%{\kern -1cm}$ \\
  ${\kern -1cm} 20\%$ & $ 20^{+15}_{-14}\%{\kern -1cm}$ \\
  ${\kern -1cm} 30\%$ & $ 30^{+15}_{-14}\%{\kern -1cm}$ \\
  ${\kern -1cm} 40\%$ & $ 40^{+14}_{-16}\%{\kern -1cm}$ \\
  ${\kern -1cm} 50\%$ & $ 50^{+15}_{-15}\%{\kern -1cm}$ \\
  ${\kern -1cm} 60\%$ & $ 60^{+16}_{-14}\%{\kern -1cm}$ \\
  ${\kern -1cm} 70\%$ & $ 70^{+16}_{-14}\%{\kern -1cm}$ \\
  ${\kern -1cm} 80\%$ & $ 80^{+16}_{-15}\%{\kern -1cm}$ \\
  ${\kern -1cm} 90\%$ & $ 90^{+10}_{-15}\%{\kern -1cm}$ \\
  ${\kern -1cm}100\%$ & $ 100^{+0}_{-16}\%{\kern -1cm}$ \\
  \enddata
  \tablecomments{For each distribution from which we draw an
    observed sample (given in the first column and specified by the
    fraction of aligned systems in the distribution), we show the
    derived underlying distribution (see \refsecl{constraints_phot}
    and \reffig{ks_phot} for details). The population of \sini\
    measurements expected from the pre-selected TESS targets
    accurately recovers the correct distribution.\vspace{-0.76cm}}
\end{deluxetable}

\subsection{A More Conservative Simulation}
\label{sec:conservative}

As described in \refsecl{errors}, the simulations presented to this
point rely upon a number of assumptions about our understanding of
stellar populations, achievable measurement precisions, and the power
of analysis techniques that may in some cases be described as
optimistic. In this section, we identify two of the more optimistic
assumptions that we have made and we describe and summarize a second
set of simulations in which we temper those assumptions and present a
more conservative outlook for the application of IPR to TESS planetary
systems.

At the heart of the technique we have outlined is the assumption that
we can generate realistic \sini\ distributions for the TESS stellar
population. If our simulated stellar properties do not reflect
reality, we may be introducing a systematic bias to our final
results. In this context, we may worry about applying the
gyrochronology relation of \refeql{4} to stars outside of its intended
color range. In particular, we apply this relation to generate
rotation periods of late-type stars, but as discussed in
\refsecl{sample}, stars cooler than about $5000$\ K are not expected
to obey this relation. While this assumption leads to conservative
rotation period errors, it almost certainly also leads to a slightly
unrealistic distribution of rotation periods, which may bias final
results. In our conservative simulations, we therefore remove from the
sample all host stars cooler than $5000$\ K. While these stars are
numerous within the TESS parent sample, our previously imposed cuts
requiring $\vsini > 2\ \kms$\ and $\prot < 14$\ days had already
removed the vast majority of late-type stars. Nevertheless, because
the fraction of systems displaying detectable photometric rotational
modulation is much larger for late-type stars, they did represent
about $35\%$\ of the systems for which we expected to measure
\sini. These conservative simulations therefore predict only about
$130$\ (instead of $\mysim 200$) measured \sini\ values.

In addition to limiting the number of stars we include, we also
consider the measurement uncertainties we assign to our
observations. For example, we have assumed that TESS rotation periods
will be determined as precisely as \kep\ rotation periods. While the
typical photon noise should be similar, TESS targets will only be
observed for $27.4$\ days per pointing, so they will not have the
benefit of years-long observations. To estimate the TESS rotation
period precisions, we simulate TESS photometric data sets with white
noise of $200$\ ppm, a cadence of $2$\ minutes, and a duration of
$27.4$\ days. We vary spot latitudes and amplitudes, and stellar
inclinations and rotation periods out to $14$\ days, and we attempt to
measure rotation periods by fitting the peaks of the autocorrelation
function \citep[see][]{mcquillan:2014}. We confirm that for the
typical photometric amplitudes observed by \kep, we can recover the
correct period, and we calculate a measurement precision that depends
almost entirely on the rotation period itself \citep[as was found
  by][]{mcquillan:2014}. Our simple light curve simulations do not
consider multiple spot groupings, potential instrumental systematics,
or contamination from other stars in the aperture, so errors may be
somewhat larger than our analysis predicts. There are also factors
that may {\em improve} precisions (e.g., a more sophisticated
analysis, the contribution of ground-based follow-up, or the fact that
a large number of planet hosts will actually lie in fields observed by
TESS for as long as a year), but we nevertheless inflate our
calculated precisions by a factor of four in an attempt to
qualitatively account for additional sources of uncertainty in
\prot. The result is that we adopt $\sigma_\prot$\ approaching
$5\%$\ for rotation periods near $14$\ days (compared to $\prot 
\lesssim 1\%$\ for similar stars in the \kep\ sample).

Using these revised sample selection criteria and \prot\ errors, we
repeat the experiment described in previous sections to determine the
precision with which we can differentiate the underlying obliquity
distribution using a set of observed \sini\ measurements. Despite the
reduced sample size and precisions with which we can determine \prot,
we again find promising results: we estimate that even in these more
conservative simulations, $f_{\rm al}$\ can be determined to within
better than $0.2$. How such a promising result is obtained is not
immediately obvious, but can be explained fairly easily. While our
sample was reduced in size and individual measurement precision, the
stars that were ultimately lost from the sample provided the least
information about the underlying distribution. These cool stars are
the ones with the slowest rotational velocities, and consequently the
largest fractional uncertainties on their \vsini\ measurements and the
most poorly determined \sini\ values. Their inclusion in the observed
\sini\ distribution therefore improves the counting statistics but
smears the shape, both of which are used to distinguish between
simulated distributions. As a result, removing a large number of them
from the sample, as we have done in these conservative simulations,
only slightly reduces the information in the \sini\ distribution.

\section{Applications and Improvements}
\label{sec:apps}

While we have demonstrated the value of the IPR technique using
precise Gaia distances and angular diameters (rather than estimated
linear radii) for TESS transiting planets, it can easily be applied to
other surveys and other types of systems. For example, \kep\ systems
can be examined almost immediately upon release of Gaia distances, as
the planets and rotation periods have already been identified (with
occurrence and detection rates well-understood), \vsini\ measurements
have already been made, and broadband photometry exists for most stars
in the field. Similarly, as the two-wheeled \kep\ mission (K2) continues
to survey the ecliptic plane, the requisite data should become
available. Furthermore, the all-sky TESS FFIs should enable detection
of {\em more} planets than will observations of the pre-selected
target stars (see \sull). Observations will be more difficult for these
fainter stars, and measurement errors will be larger, but these
samples should eventually contribute significantly to constraints on
host star obliquities.

One caveat is that to apply IPR in the manner we have presented---to
constrain the underlying obliquity distribution---one needs to
understand the target sample and detection probabilities well enough
to simulate an accurate stellar and planetary population. In addition,
a large observing program would need to be undertaken to obtain high
signal-to-noise, high resolution spectra of the planet hosts with
known rotation periods in order to measure \vsini\ (and \teff\ for
calculating $\Theta$ via $F_{\rm bol}$). Much of this data already
exists for \kep\ targets (and now K2 targets), and has been provided
by the \kep\ team and members of the broader exoplanet community as
part of the Community Follow-up Observing Program
(CFOP\footnote{\url{https://cfop.ipac.caltech.edu}}), but additional
uncertainties may be introduced by the multiple instruments used and
widely varying quality of the spectra. While another community effort
will be most likely to provide the requisite data, an observing
program dedicated to the task would ideally be carried out on a single
suitable instrument. We note here that while the TESS target list and
expected number of planet hosts are daunting, these observations would
only need to be carried out for the $\mysim 200$\ systems amenable to
this analysis.

As for applying IPR to individual systems, the sky-plane projected
obliquities have been measured for many hot Jupiters via the
Rossiter--McLaughlin effect, and combination with the line-of-sight
projection via IPR should, in some cases, provide the true obliquity,
especially when using precise Gaia distances and angular diameters
rather than radius estimates from stellar models.

It should also be noted that the orientation of angular momenta is an
interesting diagnostic not only for short-period transiting
planets. The same method can be used for eclipsing binaries, or even
for non-transiting planets or debris disk systems. Directly imaged
planet hosts, for example, tend to be bright, young, large, and
nearby, all of which improve the measurements needed for IPR: young
stars tend to rotate more rapidly (leading to precise \vsini) and with
larger amplitudes (leading to easy \prot\ detection), and their
brightness, size, and proximity make them easily resolvable. While the
orientation of the planet or disk is not limited to an edge-on
orientation as is the case for transiting systems, if a planet shows
an orbital arc, its orbital inclination can be constrained, and one
could therefore probe angular momenta in planetary populations at wide
separations. Similarly, debris disk orientation in relation to the
stellar spin axis can constrain the primordial alignment in planetary
systems.

Though the potential applications of IPR are exciting, the biggest
hurdles will be improving measurement precision (thereby improving
individual system obliquity determinations as well as the power of
population analyses) and eliminating systematic uncertainties that
might bias our results. In most cases, \vsini\ is the limiting factor
because slow rotation is difficult to measure precisely. Using
ultra-high resolution spectrographs could reduce the error
contributions from \vsini\ (although this would increase observing
costs), as could analysis techniques developed specifically for
precise \vsini\ determination, which has not typically been a goal in
the exoplanet literature. Such improvements may also justify inclusion
of more slowly rotating stars in our sample, perhaps down to $1.5$\ or
$1.0$\ \kms, leading to a larger sample of obliquities around a wider
range of spectral types, and therefore a more powerful, less biased
population analysis. As we also demonstrated in the conservative
simulations of \refsecl{conservative}, the precision of the
measurements is as important as the sample size, and because
\vsini\ is the limiting factor for many systems---particularly
low-mass hosts---improvements to \vsini\ measurements would have a
large effect on the power of IPR and would enable a study of
obliquities across a wider range of stellar masses. Making use of
extremely precise distances and color-angular diameter relations,
stellar radius measurements can already be delivered to within
$\mysim 5\%$, but using the full energy distribution to calculate the
bolometric flux should in principle be better---careful development of
this technique should also improve IPR. For TESS stars, rotation
period measurements will be limited in most pointings to $<30$\ days
(and will not be precisely determined unless multiple periods are
observed), but rotational variables can be identified and followed up
from the ground since the stars will be so bright. This can again
expand the sample for improved analysis and improve the
\prot\ precisions.

As measurement precisions increase, the method presented herein to
constrain the true obliquity distribution of exoplanets will likely be
limited by systematic uncertainties in the assumptions made while
simulating \sini\ distributions (e.g., that gyrochronology can predict
rotation periods, or that macroturbulence can be accounted for
correctly using \teff--\logg--\vmac relations). That is, we have
compared simulated observed and expected measurements {\em under the
same assumptions}. If our assumptions contain a systematic bias,
this is removed by comparing it against itself, but an analysis that
compares against real observations will inherit that bias. In
particular, differential rotation, macroturbulence, and stellar
rotational evolution remain complicating (potentially systematic)
factors. If typical spot latitudes and differential rotation strength
can be well-characterized as a function of stellar properties, errors
associated with rotation can be mitigated. Similarly, if
characterization of macroturbulence as a function of stellar
properties continues to improve, we can eliminate additional
systematics in \vsini\ measurements. The study of gyrochronology (and
departures from standard spin-down laws) remains a very active field,
further enabled by the same space missions with which we search for
planets. Continued improvements in these areas of stellar astrophysics
will benefit many areas of study, including that of planetary system
architectures.\vspace{-0.1cm}

\section{Summary}
\label{sec:summary}

Measurements of the obliquities of the hosts of short-period giant
planets have proven invaluable to the interpretation of their orbital
evolution---the wide range of obliquities that has been observed is
suggestive of an active dynamical past for the migrating planets
and/or primordial planet-forming system. Initial results for small
planets suggest that these systems are more likely than their larger
counterparts to be flat, perhaps suggestive of a different dominant
migration mechanism or in situ formation. However, comparatively
little is known about the spin-orbit alignment for small planets---a
large sample of obliquities for small planet hosts could lend strong
evidence to the debate over in situ formation, the prevalence of
primordially tilted disks, and the importance of wide companions in
producing significant obliquity. We have presented a method to obtain
such a sample by measuring the stellar inclinations of TESS planet
hosts. To do so, we take advantage of an improvement in radius
measurements made possible by precise distances that will be released
by Gaia and angular diameters measured directly via interferometry or
estimated from broadband photometry. Our simulations indicate that an
analysis like the one presented here should be able to differentiate
relatively finely between possible underlying obliquity
distributions. In the example provided---in which the underlying
distribution is modeled as a mixture of a fraction $f_{\rm
al}$\ aligned systems and ($1 - f_{\rm al}$) randomly oriented
systems---we expect an uncertainty in the derived aligned fraction of
only $\sigma_{f_{\rm al}} \mysim 0.15$\ ($\mysim 0.20$\ under more
conservative assumptions).

This technique will also be applicable to \kep\ planets, results from
K2 and ground-based surveys, eclipsing binaries, and directly imaged
planets and disks, though it will be most useful as a population
analysis for large samples. Finally, we point out that the power of
this method can be improved by making precise measurements of \vsini,
by furthering our understanding of stellar rotation and
macroturbulence, and by carefully calibrating angular diameters
derived from photometry.

\acknowledgments 
We thank Peter Sullivan for providing the simulated
TESS planet catalog, and Josh Winn, Tim Morton, Josh Pepper, Dave
Latham, Carlos Allende Prieto, Alessandro Sozzetti, and our 
anonymous referee for helpful discussion. SQ was supported by the NSF 
Graduate Research Fellowship, Grant DGE-1051030. RW acknowledges 
support from NSF AAG grant 1009634 and NASA Origins of Solar
Systems grant NNX11AC32G.

\bibliographystyle{apj}
\bibliography{master}

\end{document}